\documentstyle[twocolumn,aps,epsf]{revtex}       
\begin{document}                          	     
\draft                                    	     
\twocolumn[\hsize\textwidth\columnwidth\hsize\csname 
@twocolumnfalse\endcsname

\title{Efficient Computation of Dendritic Microstructures using Adaptive Mesh Refinement}
\author{Nikolas Provatas$^{1,2}$, Nigel Goldenfeld$^{1}$, and Jonathan Dantzig$^{2}$ }

\address{
$^1$University of Illinois at Urbana-Champaign, Department of Physics
\\1110 West Green Street, Urbana, IL, 61801
}

\address{
$^2$  University of Illinois at Urbana-Champaign, Department of Mechanical and 
Industrial
Engineering 
\\1206 West Green Street, Urbana, IL, 61801
}

\date{\today}

\maketitle

\begin{abstract}
We study dendritic microstructure evolution using 
an adaptive grid, finite element  method applied to
a phase-field model. The computational complexity of our
algorithm, per unit time, scales linearly with system size, rather than the
quadratic variation given by standard uniform mesh schemes.
Time-dependent calculations in two dimensions are in good
agreement with the predictions of solvability theory, and can
be extended to three dimensions and small undercoolings.

\end{abstract}

\pacs{05.70.Ln, 81.30.Fb, 64.70.Dv, 81.10.Aj}
]

Dendrites are the primary component of solidification 
microstructures in metals. The formation, shape, 
speed and size of dendritic microstructures has been a topic of intense 
study in the past 10-15 years.  Experiments \cite{Hua81,Gli84} by
Glicksman and coworkers on succinonitrile (SCN) and other transparent
analogues of metals have been accurate enough to provide tests of theories
of dendritic growth, and have stimulated considerable 
theoretical progress\cite{Lan80,LanII80,Kes88}.
The experiments have clearly demonstrated that 
naturally growing dendrites possess a unique steady state tip, characterized by
its velocity, radius of curvature and shape, which leads the to time-dependent 
sidebranched dendrite as it propagates.  

The earliest theories of dendritic growth solved for the diffusion field
around a self-similar body of revolution propagating at constant speed
\cite{Iva47,Hor61}. In these studies the diffusion field determines
the product of the dendrite velocity and tip radius, but neither quantity
by itself. Adding capillarity effects to the theory predicts a unique
maximum growth speed, \cite{Tem60} but experiments showed that this
point does not represent the operating state for real dendrites.

The goal of contemporary research has been to predict 
steady state features of dendritic growth and to compute time-dependent
microstructures from numerical solutions of the equations of motion.
The purpose of this letter is to present a computationally efficient
method for time-dependent calculations, and to verify that the steady
state properties are in excellent agreement with those predicted by 
analysis of the steady state problem.

Insight into the steady state dendrite problem was first obtained from 
local models 
\cite{Lan80,LanII80,Bro83,Ben83,Ben84,Bar87,Lan86,Bre91,KessI84,KessII85,Ben93,Bre93}
describing the evolution of the interface, and incorporating the features
of the bulk phases into the governing equation of motion for the 
interface. These models were the first\cite{Ben83} to show that 
a nonzero dendrite velocity is obtained only if a source of anisotropy
 -- for example, anisotropic interface energy -- is present in the 
description of dendritic evolution. 
Subsequently, it was shown that the spectrum of allowed steady state velocities
is discrete, rather than continuous,
and the role of anisotropy was understood theoretically,
both in the local models and the full moving boundary problem 
\cite{LanII80,Kes88,Bre91,Pom91}.
Moreover, only the fastest of a spectrum of steady state velocities 
is stable, thus forming the operating state of the dendrite.  It is widely believed
that sidebranching is generated by thermal or other statistical fluctuations on
a microscopic scale, which are amplified by advective diffusion.
This body of theoretical work is generally known as solvability
theory.

Numerically solving the time-dependent Stefan problem, or variations of it, 
is difficult, requiring front tracking and lattice deformation to contain the 
interface at predefined locations on the grid \cite{Alm91}. These difficulties have 
been partially addressed
by the introduction of the {\it phase-field} model, which introduces an auxiliary
continuous order parameter $\phi ({\bf r})$ that couples to the evolution 
of the thermal field. The phase field interpolates between the solid
and liquid phases, attaining two different constant values in either phase, with
a rapid transition region in the vicinity of the solidification front.  
The level set of $\phi ({\bf r})= 0$ is identified 
with the solidification front, and the dynamics of $\phi$ are carefully designed
so that the level set dynamics follows that of the evolving solidification  front
\cite{LanS86,Fix83,Col85,Hoh77,War95,Whe92,Kar94,Eld94,Whe96,Kob93,Pro96,Wan96}..

The phase-field model finesses the problem of front tracking, but it is
still prohibitively expensive for large systems.  The cost is driven by the combined
requirements of fine resolution near the interface and a domain size set by
the diffusion length and time for the system to evolve to steady state. The
grid spacing must be small enough that the phase-field model converges to
the Stefan problem, often referred to as the sharp interface limit. 
Recent work by Karma and Rappel \cite {Kar95},
involving an improved representation for the temperature field within
the interface, has extended this limit to the order of the
capillary length, typically $10^{-8}$m.  While improved asymptotics does 
help, the large system size needed to contain large diffusion-limited 
structures which form at low undercoolings, and the extended time scales 
have remained out of reach.  Unfortunately, it is precisely 
these conditions that prevail in the most successful experiments done 
to date \cite{Gli84}. 

Our contribution in this article is to show how Karma and Rappel's phase field 
model can be implemented in a computationally efficient manner, thus
removing a significant obstacle to the numerical solution of 
large-scale solidification problems.  We exploit the fact that the 
phase and temperature fields are both essentially
constant over most of space, with significant variation only near the solidification
front itself.  This suggests that an adaptive mesh method which concentrates
grid points in the interface region will be efficient.  However, the implementation
is non-trivial, and we have found it effective to use finite element methods,
as described below.

We first illustrate that our method allows faster and 
more efficient numerical integration of phase-field models, especially  
in large systems and integration for long times.  We then examine the 
physics of dendritic growth using a phase-field model solved by this method. 
In particular, we present the convergence properties and velocity selection 
of dendrites. We also measure the effective anisotropy 
introduced by our adapting grids, and compare it to that obtained using 
uniform grid methods. 
 
We model solidification using the phase-field model used by
Karma and Rappel \cite{Kar95}. We rescale temperature $T$ by
$U=c_P(T-T_M)/L$, where $c_P$ is the specific heat at constant pressure,
$L$ is the latent heat of fusion and $T_M$ is the melting temperature. 
The order parameter is defined by $\phi$, with $\phi=1$ in the solid,
$\phi=-1$ in the liquid. The interface is defined by $\phi=0$. We rescale 
time by $\tau_o$, a time characterizing atomic movement in the interface, 
and length by $W_o$, a length characterizing the  narrow region between 
liquid and solid. The model is given by
\begin{eqnarray}
&&\frac{\partial U}{dt} = D \nabla^2 U + \frac{1}{2} \frac{\partial 
\phi}{\partial t}
\label{phase-field}\\
\nonumber
&A^2(\vec{n})&  \frac{\partial \phi}{dt} = \nabla \cdot 
(A^2(\vec{n}) \nabla \phi )  +  
(\phi - \lambda U (1-\phi^2))(1- \phi^2) \\
\nonumber
& + &
\frac{\partial }{\partial x} \left( |\nabla \phi|^2 A(\vec{n}) 
\frac{\partial A(\vec{n})}{\partial \phi_{,x}} \right) +
\frac{\partial }{\partial y} \left( |\nabla \phi|^2 A(\vec{n}) 
\frac{\partial A(\vec{n})}{\partial \phi_{,y}} \right),
\end{eqnarray}
where $D=\alpha \tau_o/W_o^2$ and $\alpha$ is the thermal diffusivity,
and where $\lambda$ is a parameter that
controls the coupling of $U$ and $\phi$.  Anisotropy has been
introduced in Eqs.~(\ref{phase-field}) by defining the width of the interface to be
$W(\vec{n})=W_o A(\vec{n})$
and the characteristic time by $\tau(\vec{n})=\tau_o A^2(\vec{n})$ 
\cite{Kar95}, with $A(\vec{n}) \in [0,1]$, and 
\begin{equation}
A(\vec{n}) = (1- 3 \epsilon) \left[  1 + \frac{4 \epsilon }{ 1 - 3 \epsilon}
\frac{(\phi_{,x})^4 + (\phi_{,y})^4 }{| \nabla \phi|^4}\right].
\label{width}
\end{equation}
The vector $\vec{n}$ in Eq.~(2) is the normal to the contours of $\phi$, 
and $\phi_{,x}$ and $\phi_{,y}$ represent partial derivatives with respect to $x$ and $y$.
The constant $\epsilon$ parameterizes the deviation of $W(\vec{n})$ from $W_o$.

Karma and Rappel \cite{Kar95} derived asymptotic relationships between 
the parameters of Eqs.~(\ref{phase-field}) which allow them to operate 
in the sharp interface limit, where $U$ at the interface satisfies
$U_{\rm int} = -d(\vec{n}) \kappa - \beta(\vec{n}) V_n $,
where $d(\vec{n})$ is the capillary length, $\kappa$ is the local curvature, 
$\beta(\vec{n})$ is the interface attachment kinetic coefficient 
and $V_n$ the normal speed of the interface, all assumed in dimensionless form 
here.  In terms of Eq.~(\ref{width}), 
$d(\vec{n}) = d_o \left[ A(\vec{n}) + \partial_{\theta}^2 A(\vec{n}) \right]$, 
where $d_o=E_1/\lambda$, where $E_1=0.8839$ \cite{Kar95} and $\theta$ 
is the angle between $\vec{n}$ and the $x$-axis.
Karma and Rappel \cite{Kar95} also showed that $W_o$, $\tau_o$ and $\lambda$ can be 
chosen so as to simulate arbitrary values of $\beta$.
In particular, choosing $\lambda=C_1 D =C_1 \alpha \tau_o/W_o^2$, with  $C_1=1.5957$, 
makes $\beta=0$, a limit which is appropriate for SCN \cite{Kar95}. 

We compute four-fold symmetric dendrites in a quarter-infinite space,
initiated by a small quarter disk of radius $R_o$ centered at the origin. 
The order parameter is initially set to its equilibrium value 
$\phi_o(\vec{x})=-\tanh((|\vec{x}|-R_o)/\sqrt{2} )$ 
along the interface.  The temperature is initialized to be everywhere
equal to its far-field undercooling
$U(|\vec{x}|)=c_p (T_\infty -T_m)/L = -\Delta$.

We simulate Eqs.~(\ref{phase-field}) on an adaptive grid of 
linear isoparametric quadrilateral and triangular finite elements, 
formulated using Galerkin's method. Element data are arranged on a 
two dimensional element-quadtree data structure \cite{She88},
making our code scalable. The grid is locally refined to have 
a higher density of elements in the vicinity of the
interface of the $\phi$-field, as well as in an extended region
in the liquid which contains the  $U$-field. The criterion
for refinement  is based on changes in fluxes of both the $\phi$ and
$U$ fields. Typically, the grid is adapted every $100$ time steps
which permits the $\phi$ and $U$ fields to remain 
within the refined range between regridding updates.
We allow a difference of at most one level of refinement between neighboring
quadrilateral elements. In such a case the quadrilateral element of lower 
level of refinement has an extra side node. 
The extra nodes are resolved 
with triangular elements.  Details of our algorithm will be presented in an upcoming publication.

When using an adaptive grid procedure, the concept of a grid spacing is
replaced by that of a minimum grid spacing $\Delta x_{\rm min}$, representing the
finest level of spatial resolution. We found that best convergence is obtained
when the algorithm {\it layers} the grid so the highest density of 
elements appears around the $\phi$ interface, whose width is of 
order $1$.  The $U$-field ahead
of the $\phi$ field is of order $D/V_n$ and has smaller gradients than 
$\phi$, and so it is encompassed by a uniform mesh of grid spacing 
$2 \Delta x_{\rm min}$ and $4 \Delta x_{\rm min}$.
We found that the convergence of our solutions is relatively insensitive
to $\Delta x_{\rm min}$. For a test case of dendrites grown at $\Delta =0.55$,
$D=2$, $\epsilon=0.05$, and integration time step $dt=0.016$,
our solutions for the steady state 
velocity converge to that given by solvability theory to within a few percent
for $0.3 \le \Delta x_{\rm min} \le 1.6$. 

\begin{figure}
\begin{center}
\leavevmode \mbox{\epsfysize=7.5cm\epsfbox{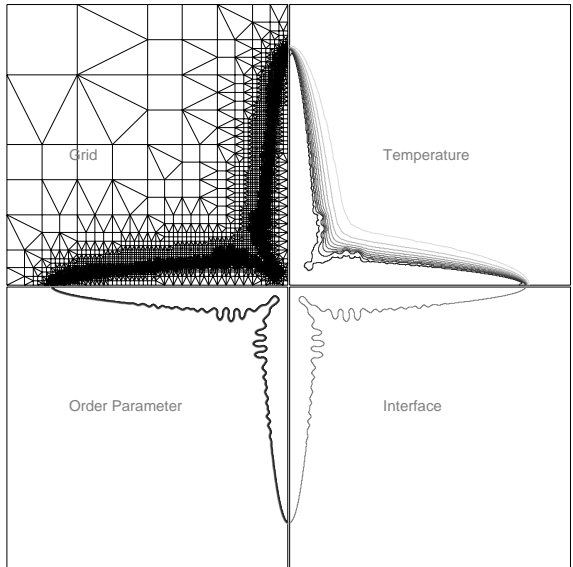}} \\
\caption{A dendrite grown using the adaptive-grid method for
$\Delta=0.55$, $D=2$, $\epsilon=0.05$, and $dt=0.016$. Clockwise,
beginning at the upper right the figures show contours of the $U$-field,
the contour $\phi=0$, contours of the $\phi$-field and
the current mesh.
}
\label{dendrite_picture}
\end{center}
\end{figure}

Fig.~\ref{dendrite_picture} shows a typical dendrite $10^5$ times steps 
into its evolution computed using our adaptive grid method. For this case,
$\Delta=0.7$, $D=2$, $\epsilon=0.05$, and the time step $dt=0.016$. 
The system size is $800 \times 800$ with $\Delta x_{\rm min}=0.8$.
Side branching 
is evident, and arises due to numerical noise.  About half the computational 
domain is shown. This calculation took approximately 10 
cpu-hours on a Sun UltraSPARC 1200E workstation.
%

\begin{figure}
\begin{center}
\leavevmode\mbox{\epsfxsize=7.5cm\epsfbox{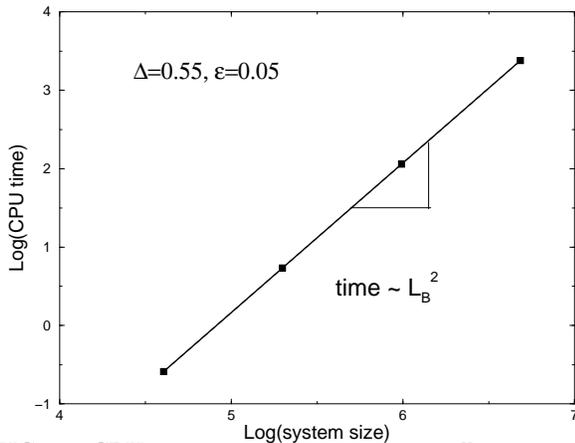}} \\
\caption{CPU time vs.~the system size, illustrating the
quadratic dependence of computing time for a dendrite to move through
the system on linear dimension $L_B$.
}
\label{cpu}
\end{center}
\end{figure}

We examined the cpu-scalability of our adaptive grid algorithm with
system size by growing dendrites in systems of linear dimension $L_B$
and measuring the cpu time $R_t^a$ for the dendrite to traverse
the entire system. For $\Delta=0.55$, $\Delta x_{\rm min}=0.4$ 
and the other parameters the same 
as those used in computing Fig.~\ref{dendrite_picture}, the relationship  
between $R_t^a$ and $L_B$ is shown in Fig.~\ref{cpu} where we see that 
$R_t^a \sim L_B^2$. The number of calculations performed, per unit time, 
is proportional to the number of elements in our grid, which is set by 
the arclength of the dendritic interface being simulated 
%
%
multiplied by the diffusion length $D/V_n$.  For a parabolic shape 
the arclength is approximately $L_B$. Thus, since the dendrite moves 
at a constant velocity $V_n$,
\begin{equation}
R_t^a = \left[ \frac{R_o^a D }{V_n^2 \Delta x_m^2} \right] L_B^2,
\label{cputime}
\end{equation}
where $R_o^a$ is a constant that depends on the implementation.
The cpu time $R_t^u$ needed to compute a full dendritic microstructure on
a uniform grid scales as $R_t^u =[R_o^u /( V_n \Delta x_m^2) ]L_B^3$.
For large system sizes, our method will be faster than uniform grid methods
by a factor $L_B$.

\begin{figure}
\begin{center}
\leavevmode \mbox{\epsfxsize=7.5cm\epsfbox{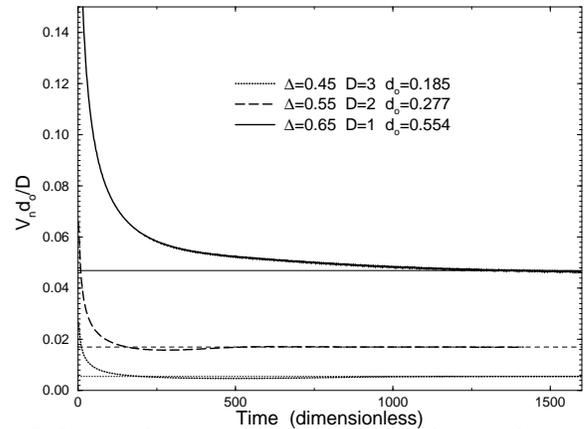}} \\
\caption{The time evolution of the dimensionless tip velocity for
undercooling $\Delta=0.45, 0.55, 0.65$. The horizontal
lines are the results of solvability theory.
}
\label{velocities}
\end{center}
\end{figure}

We tested the velocity selection mechanism of the phase-field model solved by 
our adaptive-grid algorithm for various undercoolings. 
In all cases we found very good agreement with the results of 
solvability theory. Fig.~\ref{velocities} shows the rescaled tip 
velocity $V_n d_o/D$ vs. time for three different undercoolings, 
$\Delta=0.45, 0.55, 0.65$, with corresponding dimensionless thermal 
diffusivities $D=3, 2, 1$, and dimensionless capillary length 
$d_o=0.185, 0.277, 0.554$ respectively.  
In all cases the steady state tip velocity of our dendrite reproduces solvability 
theory -- shown by the horizontal lines -- within a few percent. 
%

\begin{figure}[h]
\begin{center}
\leavevmode \mbox{\epsfxsize=7.5cm\epsfbox{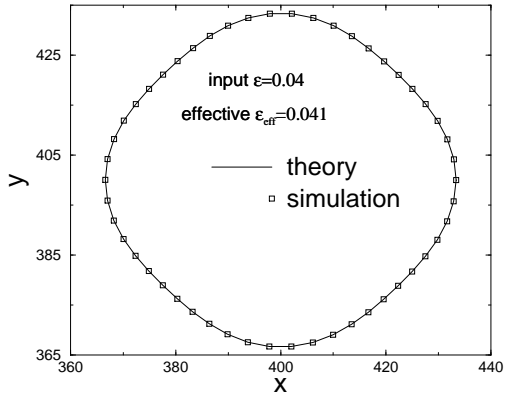}}
\caption{The equilibrium shape of the interface,
for an input anisotropy $\epsilon=0.04$.  The effective anisotropy
$\epsilon_{\rm}=0.041$.
}
\label{anisotropy}
\end{center}
\end{figure}

We also tested the effective anisotropy of our dynamically adapting lattice
following the method of Karma and Rappel\cite{Kar95}. We couple $\phi$ to 
an initially constant  background temperature $\Delta_b=d_o/R_o$. This 
maintains the interface in equilibrium for the case of isotropic surface energy. 
When growing in the presence of anisotropy, $\Delta_b$ is adjusted 
dynamically so as to maintain the velocity of the interface, 
measured along the x-axis, to zero.  Eventually our crystal neither 
shrinks nor grows, and its final equilibrium shape is fitted to the form
\begin{equation}
R(\theta) = R_o (1 + \epsilon_{\rm eff} \cos \theta )
\label{aniso}
\end{equation}
where $R$ is the equilibrium radial co-ordinate the crystal
and $\theta$ is the polar angle measured from its center. The effective 
anisotropy represents the modification to $\epsilon$ 
due to the grid.  Fig.~\ref{anisotropy} illustrates a 
crystal grown to equilibrium using an input anisotropy $\epsilon=0.04$. 
Using to Eq.~(\ref{aniso}), we find $\epsilon_{\rm eff}=0.041$. This 
accuracy is typical.
%

This letter presents a new adaptive grid algorithm that 
is used to study solidification microstructures using adaptive refinement 
of a finite element grid. Our method is used to solve
the phase-field model given by Eqs.~(\ref{phase-field}).  Our main result is that 
our solution time scales linearly with system size, rather than
quadratically as one would expect in a uniform mesh.
This allows us to solve the phase-field model in much larger systems and 
for longer simulation times. We showed that the convergence of our 
solutions remains accurate over a large range of $\Delta x_m$.
Furthermore, dendritic tip speeds were found to be reproduced 
within a few percent of the theoretical values predicted 
by solvability theory. The effective anisotropy induced by our method
was found to be a few percent of the input anisotropy.  
The speed increase of our method allows us
to investigate dendritic microstructures at undercoolings somewhat lower
than $\Delta=0.1$ in 2D. These results, as well as work in 3D will 
appear in an upcoming publication.

Acknowledgements:   This work has been supported
by NASA Microgravity Research Program, under Grant NAG8-1249.
We also thank Wouter-Jan Rappel for providing the Green's function 
steady-state code used to test our simulations.



\begin{thebibliography}{99}

\bibitem{Hua81}
S.-C. Huang and M.E. Glicksman, Acta. Metall., {\bf 29}, 701 (1981)]

\bibitem{Gli84}
M. E. Glicksman, Materials Science and Engineering, {\bf 65} (1984).

\bibitem{Lan80}
J.S. Langer, Rev. Mod. Phys. {\bf 52}, 1 (1980); 

\bibitem{LanII80}
J.S. Langer, 
``Lectures in the Theory of Pattern Formation", in {\em Chance and Matter}, Les Houches Session XLVI, edited by J. Souletie, J. Vannenimus and R. Stora (North
Holland, Amsterdam, 1987), p. 629.

\bibitem{Kes88}
D.A. Kessler, J. Koplik and H. Levine, Adv. Phys. {\bf 37}, 255 (1988).

\bibitem{Iva47}
G. P. Ivantsov, Dokl. Akad. Nauk USSR, {\bf 58}, 1113 (1947).

\bibitem{Hor61}
G. Horvay and J. W. Cahn, Acta Metall. {\bf 9}, 695 (1961).

\bibitem{Tem60}
D. E. Temkin, Dokl. Akad. Nauk SSSR {\bf 132}, 1307 (1960).

\bibitem{Bro83}
R. Brower, D. Kessler, J. Koplik, and H. Levine, Phys. Rev. Lett., {\bf 51}, 1111, (1983).

\bibitem{Ben83}
E. Ben-Jacob, N. Goldenfeld, J.S. Langer and G. Sch\"on, Phys. Rev. Lett.,
{\bf 51}, 1930 (1983).

\bibitem{Ben84}
E. Ben-Jacob, N. Goldenfeld, B.G. Kotliar, and J.S. Langer, Phys. Rev. Lett., 
{\bf 53}, 2110 (1984).

\bibitem{Bar87}
A. Barbieri, D. C. Hong, and J.S. Langer, Phys. Rev. A, {\bf 35}, 1802 (1987).
 
\bibitem{Lan86}
J.S. Langer, Phys. Rev. A, {\bf 33}, 435 (1985).

\bibitem{Bre91}
E. Brener, and V. I. Melnikov, Adv. Phys. {\bf 40}, 53 (1991).

\bibitem{KessI84}
D. Kessler, J. Koplik, and H. Levine,  Phys. Rev. A, {\bf 30}, 3161 (1984).
 
\bibitem{KessII85}
D. Kessler, J. Koplik, and H. Levine,  Phys. Rev. A, {\bf 31}, 1712 (1985).

\bibitem{Ben93}
M. Ben Amar, and E. Brener, Phys. Rev. Lett. {\bf 71}, 589 (1993).
 
\bibitem{Bre93}
E. Brener, Phys. Rev. Lett. {\bf 71}, 3653 (1993).

\bibitem{Pom91}
Y. Pomeau and M. Ben Amar, {\it Dendritic growth and related topics}, in  {\it Solids
far from equilibrium}, ed. C. Godr\`eche, (Cambridge, 1991) 365.

\bibitem{Alm91}
R. Almgren, J. Comp. Phys. {\bf 106}, 337 (1993).

\bibitem{LanS86}
J. S. Langer, in {\it Directions in Condensed Matter} (World Scientific, Singapore, 1986), 164.

\bibitem{Fix83}
G. J. Fix, in {\it Free Boundary Problems: Theory and Applications}, 
Vol. II, edited by A. Fasano and M. Primicerio (Piman, Boston, 1983), 580.

\bibitem{Col85}
J. B. Collins and H. Levine, Phys. Rev. B, {\bf 31}, 6119 (1985).


\bibitem{Hoh77}
P. C. Hohenberg, and B. I. Halperin, Rev. Mod. Phys. {\bf 49}, 435 (1977).


\bibitem{War95}
J. A. Warren and W. J. Boettinger, Acta Metall. Mater. A {\bf 43}, 689 (1995)

\bibitem{Whe92}
A. A. Wheeler, W.J. Boettinger, and G. B. McFadden Phys. Rev. A {\bf 45}, 7424 (1992)
(1996).

\bibitem{Kar94}
A. Karma, Phys. Rev. E {\bf 49}, 2245 (1994).

\bibitem{Eld94}
K. R. Elder, F. Drolet, J. M. Kosterlitz, and M. Grant, Phys. Rev. Lett. {\bf 72}, 677 (1994).

\bibitem{Whe96}
A. A. Wheeler, G. B. McFadden, and W.J. Boettinger,  Proc. Royal Soc. London A {\bf 452}, 

\bibitem{Kob93}
R. Kobayashi, Physica D {\bf 63}, 410 (1993).

\bibitem{Pro96}
N. Provatas, E. Elder, M. Grant, Phys. Rev. B,
{\bf 53}, 6263 (1996).

\bibitem{Wan96}
S-L. Wang and R. F. Sekerka, Phys. Rev. E {\bf 53}, 3760 (1996).

\bibitem{Kar95}
A. Karma, and W.-J. Rappel, Phys. Rev. E {\bf 53}, 3017 (1995);
A. Karma, and Wouter-Jan Rappel, Preprint (1997).

\bibitem{She88}
M. S. Shepard and J. Z. Zhu, Int. J. Numer. Math. Eng. {\bf 32}, 783 (1991).

\end{thebibliography}
\end{document}